%% file: sc.tex
\documentclass[3p,twocolumn]{elsarticle}
\usepackage{graphicx}
\usepackage{subfig}
\usepackage{amsmath}
\usepackage{upgreek}
\usepackage{bm}
\usepackage{booktabs}
\newcommand{\T}{\discretionary{}{}{}}

\usepackage[printonlyused]{acronym}
\usepackage{hyperref}
\hypersetup{pdfborder={0 0 0 0}}
\input myunits.sty
\input mymath.sty

\def\panda{\leavevmode{\sc\setbox0\hbox{p}\hbox to \wd0{\vbox{\hrule\vskip1.2pt\box0}}anda}}

\def\ENDING{_jpg}

\newcommand{\geant}{GEANT}
\graphicspath{ {./pics/} }

\journal{Nuclear Instruments and Methods A} 

\begin{document} 
\begin{frontmatter} 
\title{Space-Charge Effects in an
  Ungated GEM-based TPC \tnoteref{t1}} 
\author[TUM]{F.~V.~B\"ohmer\corref{c1}}
\author[TUM]{M.~Ball} 
\author[TUM]{S.~D{\o}rheim} 
\author[TUM]{C.~H\"oppner} 
\author[TUM]{B.~Ketzer} 
\author[TUM]{I.~Konorov} 
\author[TUM]{S.~Neubert}
\author[TUM]{S.~Paul} 
\author[TUM]{J.~Rauch} 
\author[TUM]{M.~Vandenbroucke} 
\address[TUM]{Technische Universit\"at M\"unchen, Physik Department
  E18, D-85748 Garching, Germany} 

\cortext[c1]{Corresponding author, email: \texttt{felix.boehmer@cern.ch}}
\tnotetext[t1]{This work has been supported by the $7^{\text{th}}$ Framework Program of the EU, the German Bundesministerium f\"ur Bildung und Forschung and the DFG Cluster of Excellence ``Origin and Structure of the Universe''.}

\begin{abstract} 
  A fundamental limit to the application of Time Projection Chambers
  (TPCs) 
  in high-rate experiments is 
  the accumulation of slowly drifting
  ions in the active gas volume, which compromises the
  homogeneity of the drift field and hence the detector resolution.
  Conventionally, this problem is overcome by the use of
  ion-gating structures. This method, however, introduces large dead
  times and restricts trigger rates to a few hundred per second. The
  ion gate can be eliminated from the setup by the use of Gas Electron
  Multiplier (GEM) foils for gas amplification, which intrinsically
  suppress the backflow of ions. This makes the continuous operation
  of a TPC at high rates feasible. 

  In this work, Monte Carlo simulations of the buildup of ion space
  charge in a GEM-based TPC and the correction of the resulting drift
  distortions are discussed, based on realistic numbers for the ion
  backflow in a triple-GEM amplification stack. A TPC in the future
  \panda~experiment at FAIR, in which $\overline{p}p$ interaction rates up
  to $2\cdot10^{7}\,\mathrm{s}^{-1}$ will be reached, serves as an
  example for the experimental environment.  The simulations show that
  space charge densities up to $65\,\fC\,\Cm^{-3}$ are reached,
  leading to electron drift distortions of up to $10\,\mm$. 
  The application of a laser calibration system to correct these
  distortions is investigated.  Based on full simulations of the
  detector physics and response, we show that
  it is possible to correct for the drift distortions and to maintain
  the good momentum resolution of the GEM-TPC.
\end{abstract} 

\end{frontmatter}
\cleardoublepage

\input{sc_introduction}
\input{sc_gemtpc}
\input{sc_scsim}

\input{sc_distortions}

\input{sc_recovery}

\input{sc_conclusion}

\bibliographystyle{h-elsevier.bst}
\bibliography{hadron,panda,detectors,mpgd}
\begin{acronym}

  \acro{MWPC}{Multi Wire Proportional Chamber}
  \acro{TPC}{Time Projection Chamber}
  \acro{HESR}{High-Energy Storage Ring}
  \acro{FAIR}{Facility for Antiproton and Ion Research}
  \acro{PANDA}[\leavevmode{\sc\setbox0\hbox{p}\hbox to
    \wd0{\vbox{\hrule\vskip1.2pt\box0}}anda}]{Anti-Proton
    Annihilations at Darmstadt }
  \acro{TUM}{Technische Universit\"at M\"unchen}
  \acro{GEM}{Gas Electron Multiplier}
  \acro{DPM}{Dual Parton Model}
    
\end{acronym}

\end{document}

%% file: sc_introduction.tex
\section{Introduction}
\label{sec:intro}
A Time Projection Chamber (TPC) \cite{Nygren:1978rx} can be regarded
as an almost 
ideal device for charged-particle tracking. A 
large number of 3-dimensional hits measured along a particle track  
(typ. $50$--$100$) eases the task of pattern recognition in a dense
environment and allows particle identification (PID) via the
measurement of specific ionization. Large solid-angle coverage combined with
very little material in the active part of the detector makes this
device very attractive for applications in which high resolution is to
be combined with small photon conversion probability and little
multiple scattering. TPCs have been successfully used as large volume
tracking devices in many particle physics experiments, e.\,g.~PEP-4
\cite{Madaras:1982cj}, TOPAZ \cite{Kamae:1986jd}, DELPHI 
\cite{Brand:1989qv}, ALEPH
\cite{Atwood:1991bp}, NA49 \cite{Fuchs:1995nu}, STAR
\cite{Ackermann:1999kc}, CERES \cite{Adamova:2008mk} and ALICE 
\cite{Alme:2010ke}.

In its standard form, a TPC consists of a large gas-filled 
cylindrical vessel sorrounding the interaction
point, placed inside a solenoidal magnetic field to measure the
momentum of charged particles \cite{Blum:08}.  The passage
of an ionizing particle through the TPC produces a trace of
electron-ion pairs.  A uniform electric field along the cylinder axis
(``drift field''), created by the two endcaps and a field cage made of
azimuthal metallic strips on the cylinder walls maintained at a
linearly decreasing potential,  
separates positive gas ions and electrons. The
ionization electrons then drift towards the readout anode located at
one endcap of the cylinder. The transverse diffusion, which can
become quite large for drift distances of the order of $1\,\m$, is
reduced by the magnetic field parallel to the drift direction. 
At the 
endcap, a plane of proportional wires is conventionally used for
avalanche multiplication of ionization electrons. The signals induced
on an arrangement of pad electrodes provide a measurement of the track
projection onto the endplate. The third coordinate of the track is
extracted from the measurement of the drift times of the ionization
electrons. The reconstruction of a track from the measured data 
requires a precise knowledge of the 
drift of electrons and hence of the electric and magnetic fields in the
chamber. 

Electrons have drift velocities of the order of several
$\Cm\,\upmu\s^{-1}$ and are thus quickly removed from the drift volume. The
drift velocity of ions, however, is three to four orders of magnitude smaller
(see Table \ref{tab_sim1} for the values used for our simulations),
leading to a slow 
buildup of space charge in the
chamber. There are two principal sources of ions in a TPC:
\begin{itemize}
\item Gas ionization by fast charged particles traversing the drift
  volume: The created ions slowly drift towards the cathode end-plate
  of the TPC.
\item Avalanche multiplication: During avalanche amplification, a
  large amount of electron-ion pairs are created, given by the
  total gain $G$, which is typically of the order of
  $10^3$ to $10^4$. Without further measures, the ions created in the
  amplification process would move back into the drift volume and lead
  to significant distortions of the electric field.
\end{itemize}
The total amount of charge accumulated in the drift volume depends on
the rate and momentum distribution of the incident particles, the 
properties of the gas, 
and the amount of ions from the avalanche region
drifting back into the drift volume. To prevent avalanche ions from
reaching the drift volume, TPCs are normally operated in a pulsed
mode, where an electrostatic gate to the readout region is opened only
when an interaction in the target has occurred, and is closed
immediately thereafter \cite{Blum:08}.
The time needed to remove
the ions as well as the switching time of the gate constitute
dead times for the experiments, which limit the trigger rates
to several hundred per second.

Modern particle physics experiments, in contrast, require high
interaction and trigger rates and little or no dead time of the
detector systems.  In order to benefit from the advantages of a TPC in
a high-rate experiment, one has to find other means of space charge
suppression. As an alternative to ion gating, the usage of \acfp*{GEM}
\cite{Sauli:97} for gas amplification has been proposed, since these
devices feature an 
intrinsic suppression of the ion back-drift
\cite{Sauli:2003yf,Bondar:2002zq}. Operating a TPC at interaction
rates which are large compared to the inverse drift time of electrons --
which is of the order of $100\,\upmu\s$ for typical drift distances of
$1\,\m$ -- means that several events will be overlapping in one drift
frame. The TPC hence acts as an ``analog track pipeline'' with signals
arriving continuously at the readout pads. Instead of an event-based,
triggered readout, the appropriate readout mode of such a device is
then a continuous electronic sampling of signals combined with an
autonomous detection of hits, which are further processed based on
their individual time stamps.  The association of these hits to tracks
and of tracks to distinct physics events (``event deconvolution'')
requires real-time tracking capabilities of the data
acquisition system. The successful development of a continuously
running TPC, though challenging, opens the possibility to benefit from
the advantages of such a detector in future high-rate experiments. 

\begin{figure}[tb]
  \centering
  \includegraphics[width=\linewidth]{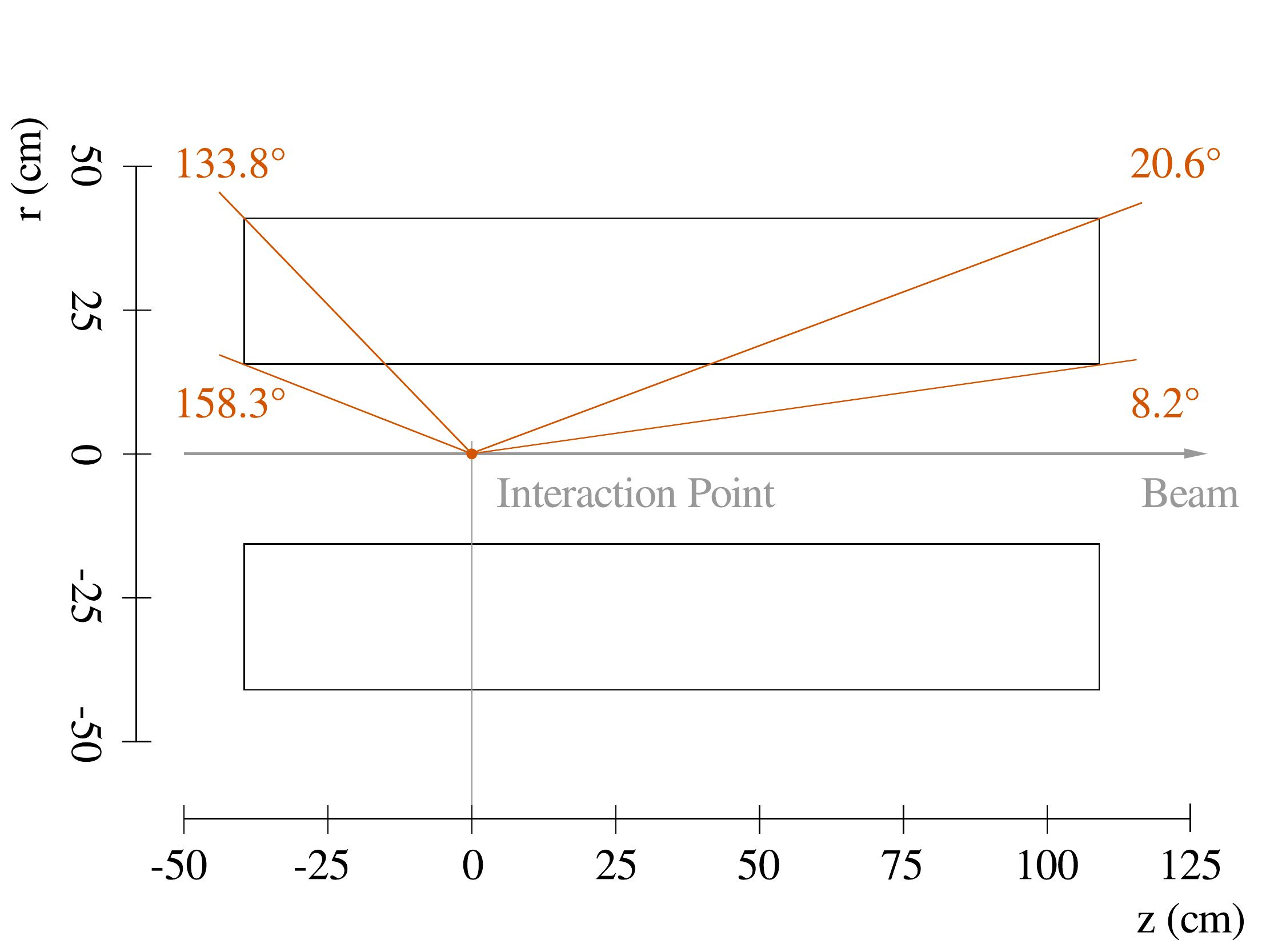}
  \caption{Geometry and dimensions of the active part of the
    \panda~TPC (cross section).} 
  \label{fig:TPC_geo}
\end{figure}

Among other options, a GEM-based TPC was proposed as the central
tracker of the experiment 
\panda~(Antiproton Annihilations at Darmstadt) 
experiment \cite{PANDA:05} at the future international Facility for
Antiproton and Ion Research (FAIR). \panda\ will use an intense,
cooled antiproton 
beam with momenta from $1.5$ to $15 \, \mathrm{GeV/}c$ impinging on
different targets and will reach luminosities up to $2 \, \cdot \,
10^{32} \, \mathrm{cm^{-2} \, s^{-1}}$, resulting in a $\bar{p}p$
interaction rate up to $2 \cdot 10^7 \mathrm{s^{-1}}$. The dimensions
of the \panda\ GEM-TPC are shown in Fig.~\ref{fig:TPC_geo}. 
It consists of  
cylindrical vessel with $150 \,\mathrm{cm}$ length and an inner
(outer) radius of $15.5 \,\mathrm{cm}$ ($41.5 \, \mathrm{cm}$) with
GEM amplification at the upstream endcap 
\cite{Weitzel:07a}, placed
in a $2\,\mathrm{T}$ solenoidal 
magnetic field. Owing to the fixed-target geometry of the
experiment, the interaction point is not located in the middle of
cylinder axis, but shifted upstream.   
The TPC was shown to provide excellent standalone 
pattern recognition and momentum resolution of the order of a few
percent \cite{Ball:2012xh}. The ungated, continuous operation mode of
the TPC at the 
envisaged event rates at \panda~ gives rise to about $4000$ tracks
which are superimposed in the drift volume at any given time.

Track densities similar to this environment are expected for the ALICE
TPC after an upgrade of the LHC, foreseen for the year 2018. At an
expected 
luminosity for Pb-Pb collisions of $6 \, \cdot \, 10^{27} \,
\mathrm{cm^{-2} \, s^{-1}}$, a continuous readout of the TPC is
required in order to make full use of the interaction rate of about
$50 \cdot 10^3 \,\text{s}^{-1}$. This cannot be achieved with the
present gated MWPC-based amplification system. A replacement by a
triple-GEM amplification using large-size foils 
is currently under evaluation.    

The ion leakage from a multi-GEM detector,
however, is considerably larger than that from a closed gating grid,
which is typically $<10^{-4}$ \cite{Alme:2010ke}. It is therefore
important 
to develop an understanding of the effects of residual space charge in
a GEM-based, continuously running TPC. To this end, we have developed
a computer 
simulation of space charge accumulation and drift distortions for a
GEM-based TPC. The
simulation is based on the following steps:
\begin{itemize}
\item transport particles from minimum bias physics events through the
detector setup and calculate their energy loss;
\item  model the drift
of ions to obtain the spatial distribution of space charge;
\item calculate the resulting electric field in the drift volume using
finite element methods;
\item solve the drift equation for electrons in
electric and magnetic fields to get a map of drift distortions as a
function of the point of generation;
\item  simulate a laser
calibration system to measure the drift distortions.
\end{itemize}

Section~\ref{sec:gem-tpc} describes the detector setup and defines the
parameters used in the simulations. In Sec.~\ref{sc_sim} we describe
how the charge accumulation is modeled to derive a space charge
distribution in the drift volume of the TPC. Section \ref{sc_dist}
details how drift distortions in the inhomogeneous drift fields are
calculated. Values for expected drift distortions in the simulated
\panda~environment are given. Finally, in Sec. \ref{sc_laser} we
demonstrate how the drift distortions can be
measured and corrected by means of an array of ionizing laser beams.

%% file: sc_gemtpc.tex
\section{Ion Backflow in a GEM-based TPC}
\label{sec:gem-tpc}

The Gas Electron Multiplier (GEM) \cite{Sauli:97} consists of a
$50\,\mum$ thin insulating Polyimide foil with Cu-coated surfaces,
typically $5\,\mum$ thick. The foil is perforated by
photo-lithographic processing, forming a dense, regular pattern of
(double-conical) holes. For standard GEM foils, the holes have an
inner diameter of $\sim 50\,\mum$ and a pitch of $140\,\upmu\m$.  The
small dimensions of the amplification structures lead to very large
field strengths $\mathcal{O}(50\,\kV\,\Cm^{-1}$) inside the holes of the GEM
foil when a moderate voltage difference of typically $300$ --
$400\,\V$ is applied between the metal layers, sufficient for
avalanche creation inside the GEM holes. 

\begin{figure}[tb]
  \centering 
  \includegraphics[width=\linewidth]{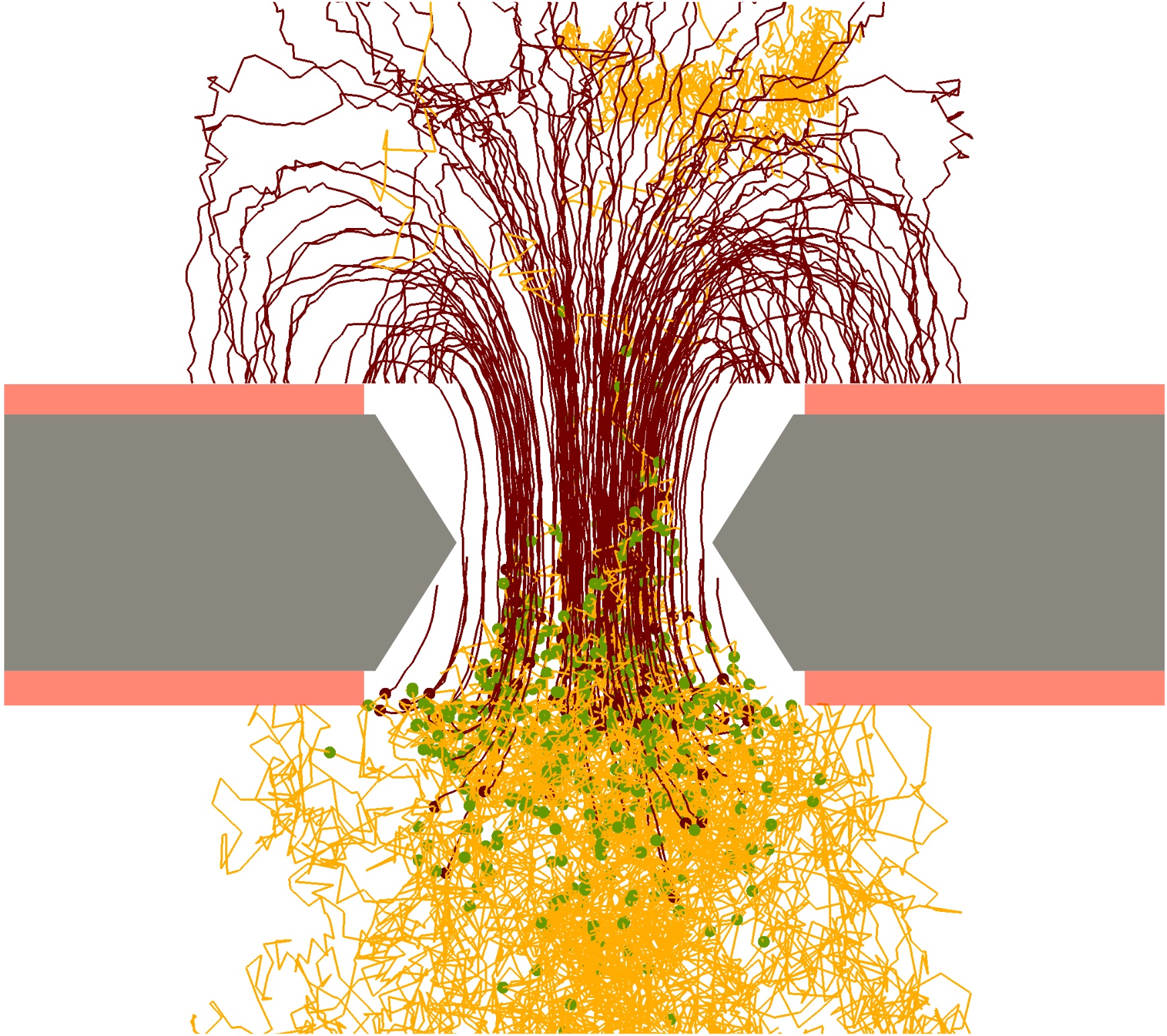}
  \caption{Garfield / Magboltz simulation of charge dynamics
    of two arriving electrons in a GEM hole. Electron paths are shown
    as light lines, ion paths as dark lines. Spots mark places where
    ionization processes have occurred. The paths have been projected
    on the cross section plane. }
  \label{fig:gem-tpc.gem_fields}
\end{figure}

The dynamics of charge movement and avalanche creation inside the GEM
holes are complicated. Figure \ref{fig:gem-tpc.gem_fields} shows a
Garfield / Magboltz \cite{GARFIELD:1984} simulation qualitatively
explaining the suppression of ion backflow from the amplification
region.  Two electrons (light lines) are guided into the GEM hole by
the drift field (here: $250\,\V\,\Cm^{-1}$), and produce avalanches by
ionizing gas molecules (dots).  The ions created in the avalanches
(dark lines) closely follow the electric field lines because of their
much smaller diffusion. Most of the ions are collected on the top side
of the GEM foil, because the field inside the GEM hole is much higher
than the field above the hole. Only a few ions drift back into the
drift volume. The extraction of electrons from the hole is facilitated
by a higher transfer field below the GEM (here: $3.75\,\kV\,\Cm^{-1}$).
The electrons can then be transferred to another amplification
stage or collected at the anode. 

Typically three GEM foils are
combined in a stack, leading to 
effective gains of the order of $10^3$ -- $10^4$ and at the same time
guaranteeing a stable operation without the occurrence of discharges
\cite{Bachmann:01e}. The effective gain $G_{\mathrm{eff}}$ of the gas
amplification system is given by the average number of electrons
arriving at the readout anode $N^-_\mathrm{A}$ divided by the number
of ionization electrons $N^-_\mathrm{I}$:
\begin{equation}
  \label{eq:effgain}
  G_{\mathrm{eff}} = \frac{N^-_{\mathrm{A}}}{N^-_{\mathrm{I}}}\quad .
\end{equation}
Gas ionization produces an equal amount of electrons and ions, hence
$N^-_{\mathrm{I}}=N^+_{\mathrm{I}}\equiv
N_\mathrm{I}$. The effective gain $G_{\mathrm{eff}}$ differs from the intrinsic gas gain
of the amplification stage, as it already includes losses due to
limited electron collection or extraction efficiencies of the
foils. These, as well as gain fluctuations, have not been explicitly
simulated in the present work.

We define the ion backflow as\footnote{We choose this quantity over
  other definitions in the literature since it can be easily measured
  via the ratio of cathode to anode current.}
\begin{equation}
  \label{eq:backf}
  I\!B = \frac{N^+_\mathrm{C}}{N^-_\mathrm{A}}\quad,
\end{equation}
where $N^+_\mathrm{C}$ denotes the number of positive ions arriving at
the drift cathode. Note that this quantity also includes a
contribution from ions created during the ionization process.  Ion
backflow values of $I\!B=0.25\%$ have been reached experimentally in a
magnetic field of $4\,\T$ in an Ar/CH$_4$/CO$_2$
(93/5/2) mixture \cite{Blatt:2006nx}.

A very convenient parameter to describe the ion backflow in
the simulation is the ratio $\epsilon$ of the number of ions drifting back
from the GEM amplification stage $N^+_{\mathrm{G}}$ and the number of
ionization electrons $N_{\mathrm{I}}$:
\begin{equation}
  \label{eq:eps}
  \epsilon = \frac{N^+_{\mathrm{G}}}{N_{\mathrm{I}}}\quad .
\end{equation}
Since the number of ions arriving at the drift cathode is given by the
sum of ions from ionization and the ones from the amplification stage,
\begin{equation}
  \label{eq:nc}
  N^+_\mathrm{C} = N_{\mathrm{I}}+ N^+_{\mathrm{G}}\quad ,
\end{equation}
the
two quantities $\epsilon$ and $I\!B$ are linked by 
\begin{equation}
  \label{eq:eps_b}
  I\!B = \frac{N_{\mathrm{I}} + \epsilon
    N_{\mathrm{I}}}{G_{\mathrm{eff}} N_{\mathrm{I}}} =
  \frac{1+\epsilon}{G_{\mathrm{eff}}}\quad .
\end{equation}

The suppression factor $\eta$ defines the efficiency of ion-backflow
suppression in the \ac*{GEM} stack:
\begin{equation}
  \label{eq:supp}
  \eta = \frac{N^+_\mathrm{G}}{N^-_\mathrm{A}} = \frac{\epsilon
    N_{\mathrm{I}}}{N_{\mathrm{I}} G_{\mathrm{eff}}} =
  \frac{\epsilon}{G_{\mathrm{eff}}}\quad ,
\end{equation}
and thus
\begin{equation}
  \label{eq:eps_fin}
  \epsilon = \eta G_{\mathrm{eff}}\quad .
\end{equation}
As a rule of thumb, a suppression factor of $\eta\sim 1/G_{\mathrm{eff}}$
is usually considered necessary in order to operate a TPC without
gating grid.

The detector gas is a crucial parameter for the performance of a TPC
\cite{Veenhof:03}.  In order to minimize multiple scattering and
photon conversion, as well as space charge effects, a popular choice
is a Ne/CO$_2$ (90/10) gas mixture \cite{Fuchs:1995nu,Alme:2010ke},
which combines low diffusion even at moderate magnetic fields, low
ionization density, large ion mobility and high radiation length.  
The results presented in this paper are based on realistic values for
the ion backflow of $I\!B=0.25\%$ and the effective gain of
$G_{\mathrm{eff}}=2000$, yielding $\epsilon=4$ according to
Eq.~(\ref{eq:eps_b}), i.e.\ four ions 
drifting back from the amplification stage per incoming electron.  The
full set of simulation parameters used for the presented results is
summarized in Table~\ref{tab_sim1}.
\begin{table}[tb]
\caption{Parameters of the space charge simulation.}
\begin{center}
\begin{tabular}[tb]{ll}
  \toprule
  Event rate & $2\cdot10^7 \, \mathrm{s^{-1}}$ \\
  Beam momentum &$p_{\bar{p}}=2.0\,$GeV/$c$ \\
  Gas mixture &$\mathrm{Ne/CO_2} \, (90/10)$ \\
  Average energy per  & $W_\mathrm{I}=36.7\,\eV$ \\
  \quad ion pair  & \\
  Nominal drift field & $400\,\mathrm{V\,cm^{-1}}$ \\
  Magnetic field & $\mathbf{B} = (0,0,2.0\,\text{T})$ \\
  Ion drift velocity & $ u^{+}=1.766\,\mathrm{cm\,ms^{-1}}$ \\
  Electron drift velocity & $ u^{-}=2.731\,\mathrm{cm\,\upmu s^{-1}}$ \\
  Longitudinal diffusion &  $ 229\,\upmu \text{m} \, \text{cm}^{-1/2}$   \\
  Transverse diffusion &  $ 128\,\upmu \text{m} \, \text{cm}^{-1/2}$  \\
  TPC dimensions  & $r=15.75$\dots$41.2\,\mathrm{cm}$ \\ \quad (active volume)& $ z=-39.5$\dots$109.5\,\mathrm{cm}$ \\
  Ion backflow factor& $\epsilon=4$ \\
  \bottomrule
\end{tabular}
\end{center}
\label{tab_sim1}
\end{table}


%% file: sc_scsim.tex
\section{Simulation of Space Charge Buildup}
\label{sc_sim}
\subsection{Ionization Charge Creation}
\label{sec:scCharge}
\begin{figure}
\centering
\includegraphics[width=\linewidth]{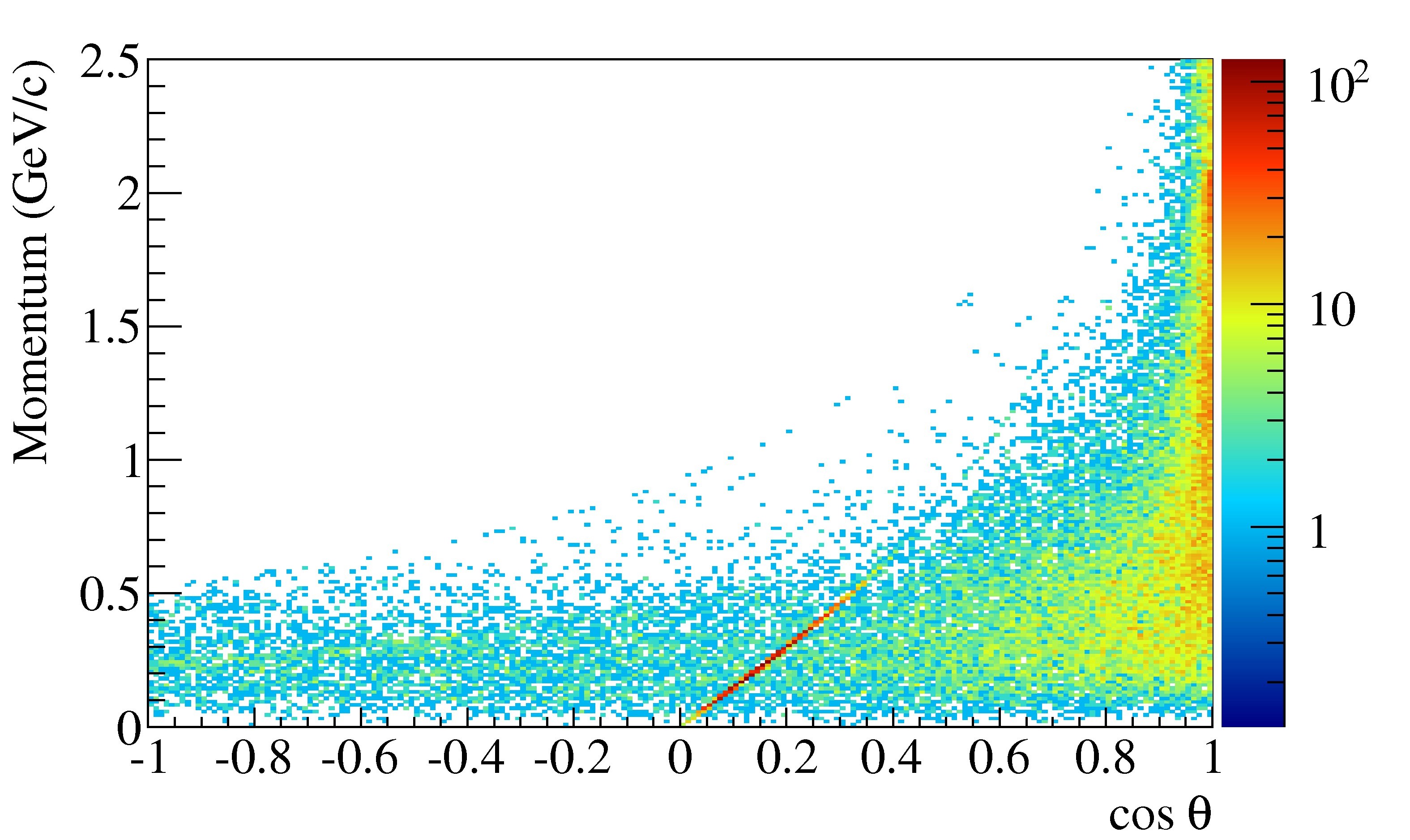}
\caption{Phase space distribution of primary particles in the
  laboratory frame produced with the DPM generator for $\overline{p}p$
  collisions at $2.0 \, \mathrm{GeV/}c$ beam momentum. The plot shows
  the absolute momentum versus the cosine of the polar angle of
  $10,000$ produced events. The sharp band of recoil protons from
  elastic scattering processes dominates the bulk of charged pions and
  kaons that are created in annihilation reactions and inelastic
  scattering.}
\label{fig:dpm}
\end{figure}

To model the spatial distribution of ionization charge in the TPC for
the case of \panda, the Dual Parton Model (DPM) \cite{DPM:94} Monte
Carlo event generator is used to simulate antiproton-proton ($\overline{p}p$)
annihilations at an incident $\overline{p}$ momentum of
$2\,\GeV/c$. The resulting phase space distribution of charged
particles is shown in Fig.~\ref{fig:dpm}.

A large sample of these generated events is then passed through a
\geant~\cite{GEANT03} simulation of the detector geometry in order to
calculate the energy loss of the particles in the TPC gas volume.  For
the simulations presented here we used GEANT3 with a simplified energy
loss model as proposed in the ALICE TPC Technical Design Report
\cite{ALICE-TPC:2000} and experimentally verified in
\cite{Christiansen:2009zza}.  From the \geant~energy deposit $\Delta
E(r,z,\phi)$ in a volume element $\diff{V}$ around a point
$(r,z,\phi)$ inside the TPC, summed up for a given number of simulated
events $N_{\mathrm{ev}}$, the number of ionization electrons/ions
$N_{\mathrm{I}}(r,z,\phi)$ produced in this volume element is
calculated using an average energy per electron--ion pair for the
Ne/CO$_2$ (90/10) gas mixture of $W_\mathrm{I}=36.7\,\eV$
\cite{Biagi:99,Veenhof:1998tt}. Here, $r$, $z$ and $\phi$ denote
cylindrical coordinates, where the $z$ axis coincides with the symmetry axis of
the TPC.  The creation rate of ionization charge in $\diff{V}$ is then
\begin{equation}
  \label{eq:geantmap}
  \dot{N}_{\mathrm{I}}(r,z,\phi,t) =
  \frac{\Delta E(r,z,\phi,t)/W_\mathrm{I}}{N_{\mathrm{ev}}} \cdot
  R(t)\quad, 
\end{equation}
with $R(t)$ being the event rate. 

\subsection{Model of Space Charge Accumulation}
\label{sec:sc_model}
In general, the charge density $\rho(r,z,\phi,t)$ at time $t$ in a
volume element $\diff{V}$ centered around a point $(r,z,\phi)$ inside the
TPC volume is given by:
\begin{equation}
\label{eqn:densint}
\rho(r,z,\phi,t) = 
\frac{e}{\diff{V}}\cdot\int_0^{t} \dot{N}(r,z,\phi,t') \,\diff{t'}\quad .
\end{equation} 
Here $\dot{N}(r,z,\phi,t')$ is the net rate of ions entering/leaving
the volume element $\diff{V}$ in the time interval
$[t',t'+\diff{t'}]$. $\dot{N}(r,z,\phi,t')$ includes both ions
directly created in gas ionization processes,
$\dot{N}_{\mathrm{I}}(r,z,\phi,t)$, as well as ions moving through the
volume surface by drift and diffusion.

Solving Eq.~(\ref{eqn:densint}) in principle requires perfect
knowledge of the dynamics of charge creation and drift/diffusion
inside the TPC volume. In order to arrive at a manageable, yet
sufficiently realistic simulation model, the following assumptions are
made:
\begin{enumerate}
\item Azimuthal symmetry: Owing to the cylindrical geometry of the
  TPC, we treat the problem in cylindrical symmetry so that the
  resulting charge density map can be represented in the
  $(r,z)$-plane.
\item Constant luminosity: We assume that $R(t)=\mathrm{const.}$,
  i.e.\ the rate of charge created in gas ionization processes is
  constant on time scales of interest to us [$\mathcal{O}(10 \,
    \mathrm{\upmu s})$]: $\dot{N}_{\mathrm{I}}(r,z,t) =
  \dot{N}_{\mathrm{I}}(r,z)$.
\item Electrostatic forces between the ions are neglected: the ion
  drift proceeds along straight lines with constant velocity $u^+$.
\item The effect of diffusion on the motion of ions is neglected. 
\end{enumerate}
In addition, we move from the infinitesimal volume elements $\diff{V}$
of Eq.~(\ref{eqn:densint}) to macroscopic volumes $\Delta V(r_i,z_j)
\equiv \Delta V^{ij}$ with constant bin widths $\Delta z =
1.0\,\text{cm}$ and $\Delta r = 0.98 \,\text{cm}$. Each such bin
($i$,$j$) then represents a ring-shaped volume in the TPC:
\begin{equation}
  \label{eq:dv}
  \Delta V^{ij} = \pi \cdot (r^2_{i,\mathrm{out}} -
  r^2_{i,\mathrm{in}}) \cdot \Delta z_j\quad ,
\end{equation}
where $r_{i,\mathrm{out}} = r_{i} + \frac{1}{2}\Delta r$ and
$r_{i,\mathrm{in}} = r_{i} - \frac{1}{2}\Delta r$ are the outer and
inner radius of bin ($i$,$j$), respectively.  The integral in
Eq.~(\ref{eqn:densint}) can then be solved numerically by a Riemann
sum:
\begin{equation}
  \label{eqn:densintDiscrete}
  \rho(r,z,t) = 
  \frac{e}{\Delta V^{ij}}\cdot\sum_{k=1}^{n} \dot{N}^{ijk} \,\Delta t 
  \equiv   \frac{e}{\Delta V^{ij}}\cdot N^{ijn}
  \quad,
\end{equation} 
with the time $t$ given by the number of steps in time bins $t =
n\cdot \Delta t$.

During each time step $\Delta t$, the ionization charge
$\dot{N}_{\mathrm{I}}^{ij}\Delta t$ is newly created in bin $(i,j)$ of
the chamber. Now the backflow of ions from the amplification stage has
to be included.  As can be seen from Table~\ref{tab_sim1}, the drift
velocity of electrons exceeds that of ions by several orders of
magnitude. In order to calculate the ion space charge we can therefore
assume instantaneous electron drift. As a consequence, a contribution
proportional to the total amount of newly added gas ionization charge
in the full TPC volume has to be added to the first bin in $z$ ($j=1$)
for each time bin $\Delta t$. The total number of ions added in each
time step is therefore
\begin{equation}
  \label{eq:backflow}
  N_{\mathrm{I}}^{ij}= \dot{N}_{\mathrm{I}}^{ij}\Delta t +
  \delta_{1,j}\epsilon \sum_{j'} \dot{N}_{\mathrm{I}}^{ij'} \Delta t\quad,
  \end{equation}
with $\delta_{i,j}$ being the Kronecker delta symbol. 

As a last ingredient, the drift of ions has to be taken into account.
For the geometry of the \panda~TPC, where the readout anode is located
at the upstream endcap of the chamber, the ion drift proceeds along
the positive $z$ direction towards the cathode endcap.  We choose
$\Delta t$ to correspond to the time needed for an ion of drift
velocity $u^{+}$ to travel the distance $\Delta z$.  This allows us to
include the ion drift in a time bin $\Delta t$ simply by shifting the
total charge by one bin in $z$.  We therefore arrive at a recursive
algorithm to calculate the space charge in the $n^{\mathrm{th}}$ time
step:
\begin{equation}
  \label{eq:densintRecursion}
  N^{ijn} = N_{\mathrm{I}}^{ij}+N^{i,j-1,n-1}\quad. 
\end{equation}

\begin{figure}[tb]
  \centering \subfloat[]  {\label{fig:fig_scdens-a}
    \includegraphics[width=\linewidth]{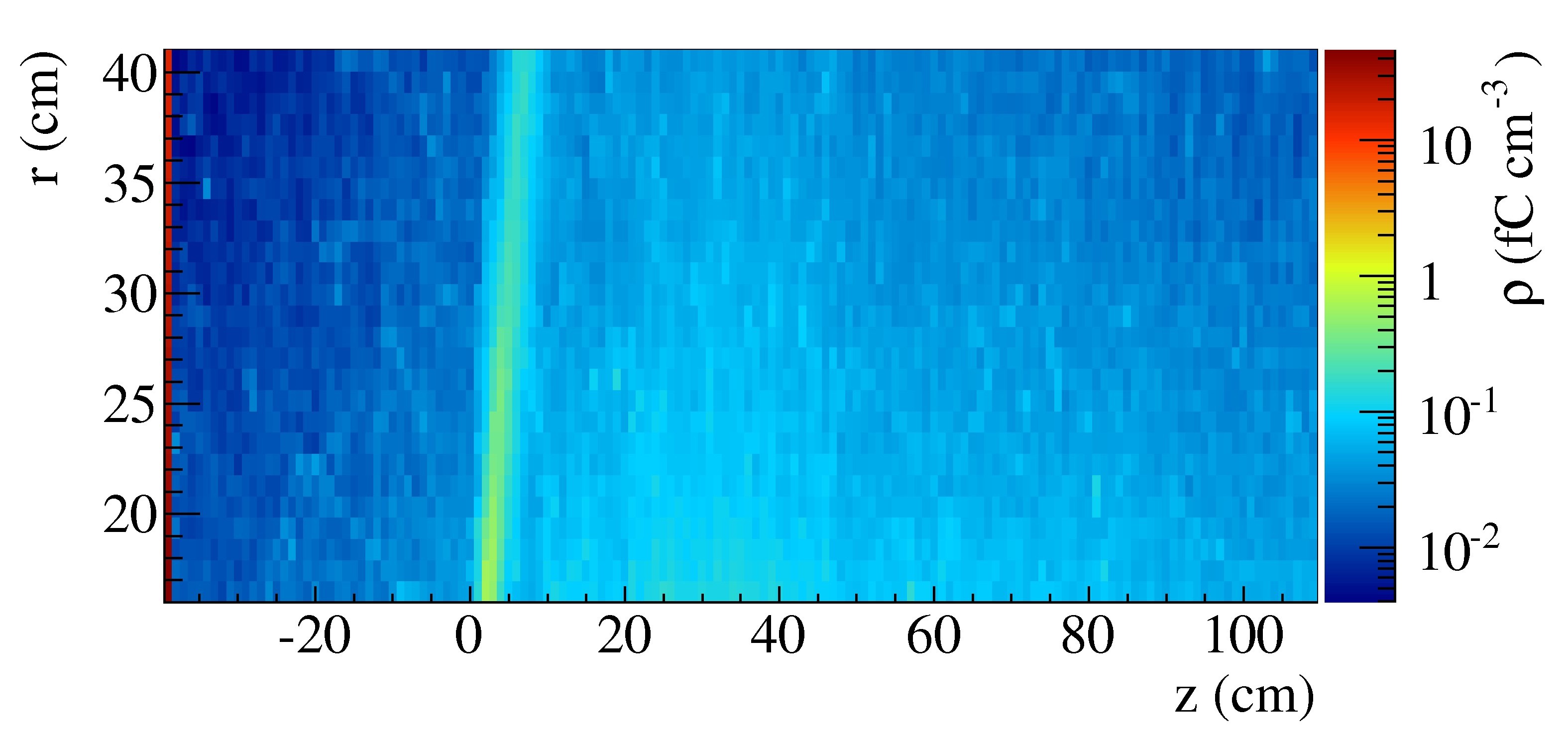}} \\
  \subfloat[]
  {\label{fig:fig_scdens-b}
    \includegraphics[width=\linewidth]{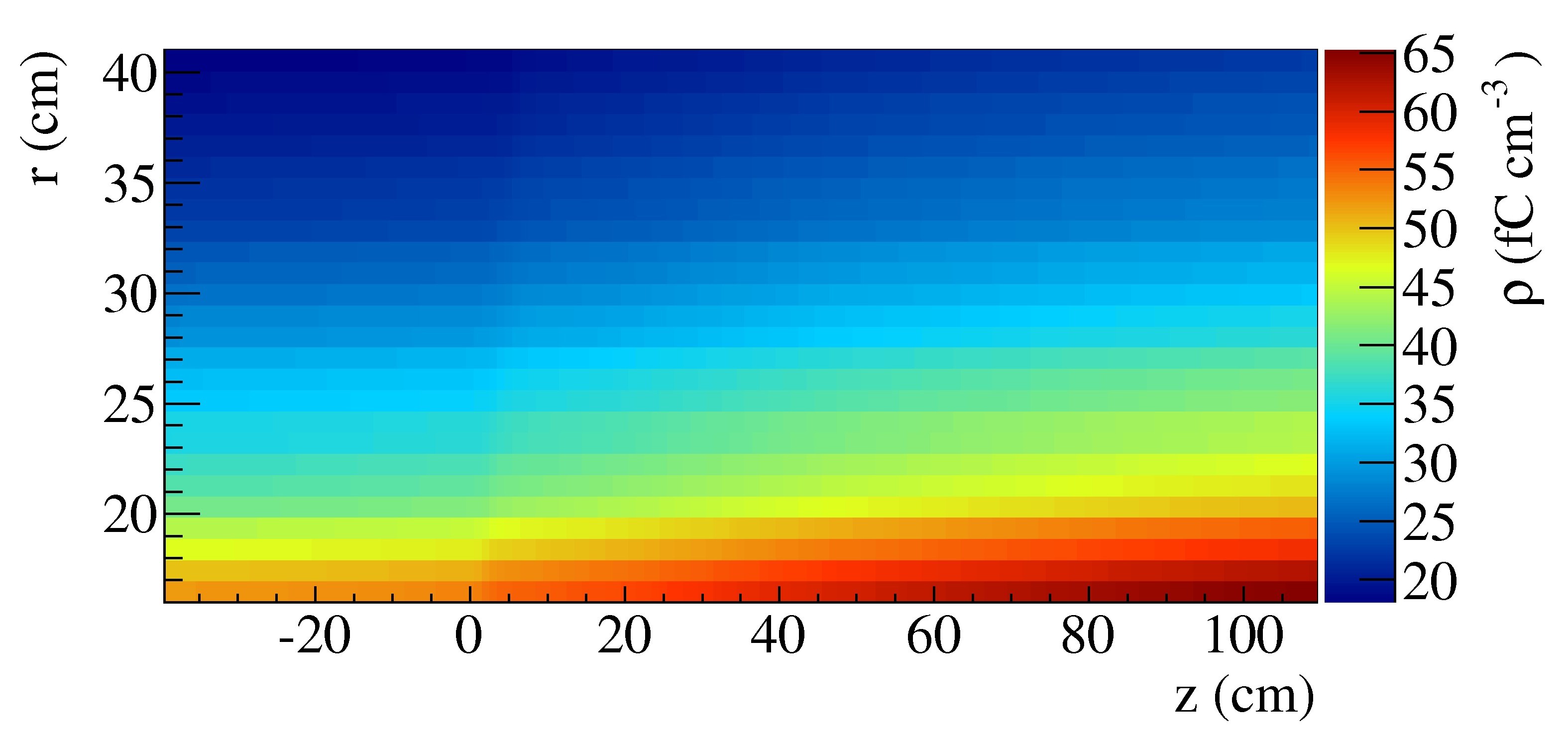}}
  \caption{Space charge density $\rho(z,r)$ in the \panda~TPC for $2\cdot10^7\,
    \mathrm{s^{-1}}$ $\overline{p}p$ annihilations at beam momentum
    $p_{\overline{p}}=2.0\,\GeV/c$ with ion back-flow factor $\epsilon=4$,
    using the recursive algorithm of
    Eq.~(\ref{eq:densintRecursion}). (a) After one step the
    distribution of ion charge from gas ionization processes is
    visible superimposed with the corresponding charge back-drifting
    from the amplification stage (red band at
    $z=-40\,\mathrm{cm}$). The band of high primary ionization density
    at $z\sim0$ is caused by slow protons from hadronic elastic
    $\overline{p}p$-scattering. (b) Final equilibrium space charge
    density.}
  \label{fig:fig_scdens}
\end{figure}

From the equations above it is clear that an equilibrium space charge
in time, $N^{ijn}=N^{ij,n-1}$, will be reached everywhere in the
chamber as soon as the perturbation due to ion backflow reaches the
cathode, i.e.\ for $t=z_{\mathrm{max}}/u^+$, where $z_{\mathrm{max}}$
is the maximum drift distance. Figure~\ref{fig:fig_scdens} shows the
space charge distributions calculated with this method after the first
step in time (i.e.\ immediately after switching on the beam) and after
reaching equilibrium.  The input parameters to the simulation are
given in Table~\ref{tab_sim1}.  After the first step,
Fig.~\ref{fig:fig_scdens-a}, the distribution of ions from direct gas
ionization by charged particles from $\overline{p}p$ annihilations
with a band at a polar angle close to $90^\circ$ from elastic
$\overline{p}p$-scattering is visible, together with the sheet of
charge at $z\approx-40\,\Cm$ starting to drift back from the
amplification region for this time bin. Due to the fixed target
geometry of \panda~and the resulting forward boost of the scattered
particles, the charge distribution is 
not symmetric in $z$ around the interaction point. 
Figure~\ref{fig:fig_scdens-b} shows the equilibrium space charge
distribution which is reached after ions from the amplification stage
have reached the cathode.  The final space charge density varies
between $20$ and $65 \, \mathrm{fC \, cm^{-3}}$. 

%% file: sc_distortions.tex
\section{Drift Distortions}
\label{sc_dist}
\subsection{Electric Field due to Space Charge}
\label{sec:distE}
From the charge-density distribution the electrostatic potential
$\varphi$ is computed by solving Poisson's equation, which in
cylindrical coordinates reads:
\begin{align}
  \label{eq:poisson}
\notag \Delta \varphi &= \frac{1}{r}\frac{\partial}{\partial
  r}\left(r\frac{\partial\varphi}{\partial r}\right) +
\frac{1}{r^2}\frac{\partial^2\varphi}{\partial\phi^2}+\frac{\partial^2\varphi}{\partial 
  z^2}   \\
  &= - \frac{1}{\epsilon_0}\rho(r,\phi,z)\quad .
\end{align}

In case of rotational symmetry, $\rho(r,\phi,z)=\rho(r,z)$, 
Eq.~(\ref{eq:poisson}) reduces to
\begin{equation}
  \label{eq:poisson_az}
  \frac{\partial^2\varphi}{\partial r^2}+\frac{1}{r}\frac{\partial\varphi}{\partial r}+\frac{\partial^2\varphi}{\partial z^2}=-\frac{1}{\epsilon_0}\rho(r,z)\quad .  
\end{equation}
Equation~(\ref{eq:poisson_az}) is solved numerically with the Finite
Element Method. The corresponding variational formulation
\cite{Grossmann:1992} is given by:
\begin{equation}
  \label{eq:var}
  \int_\Omega\Bigl[\nabla \varphi\cdot \nabla v -
  \frac{1}{r}\frac{\partial\varphi}{\partial r} v\Bigr] dz \,dr =
  \int_\Omega v\frac{\rho}{\epsilon_0} dz \,dr\quad . 
\end{equation}
Here, $\Omega$ is the drift region in the $r$-$z$-plane, $\nabla =
(\partial{}/\partial{r},\partial{}/\partial{z})$, and $v$ is an
arbitrary, dimensionless test function which satisfies the same
Dirichlet boundary condition as $\varphi$:
\begin{equation}
  \label{eq:dirichlet}
\varphi|_{\partial{\Omega}}=0, \quad v|_{\partial{\Omega}}=0\quad,
\end{equation}
where $\partial{\Omega}$ represents the boundary of the drift region,
i.e.\ the field cage wall. This boundary condition corresponds to a
perfectly conducting field cage, such that the potential of the
field-cage is fixed to zero. The software package Dolfin/FeniCs 
\cite{DOLFIN:2009} is used to realize the finite element solver.

\begin{figure}[tb]
  \centering
  \subfloat[]{\label{fig:efield-1}\includegraphics[width=\linewidth]{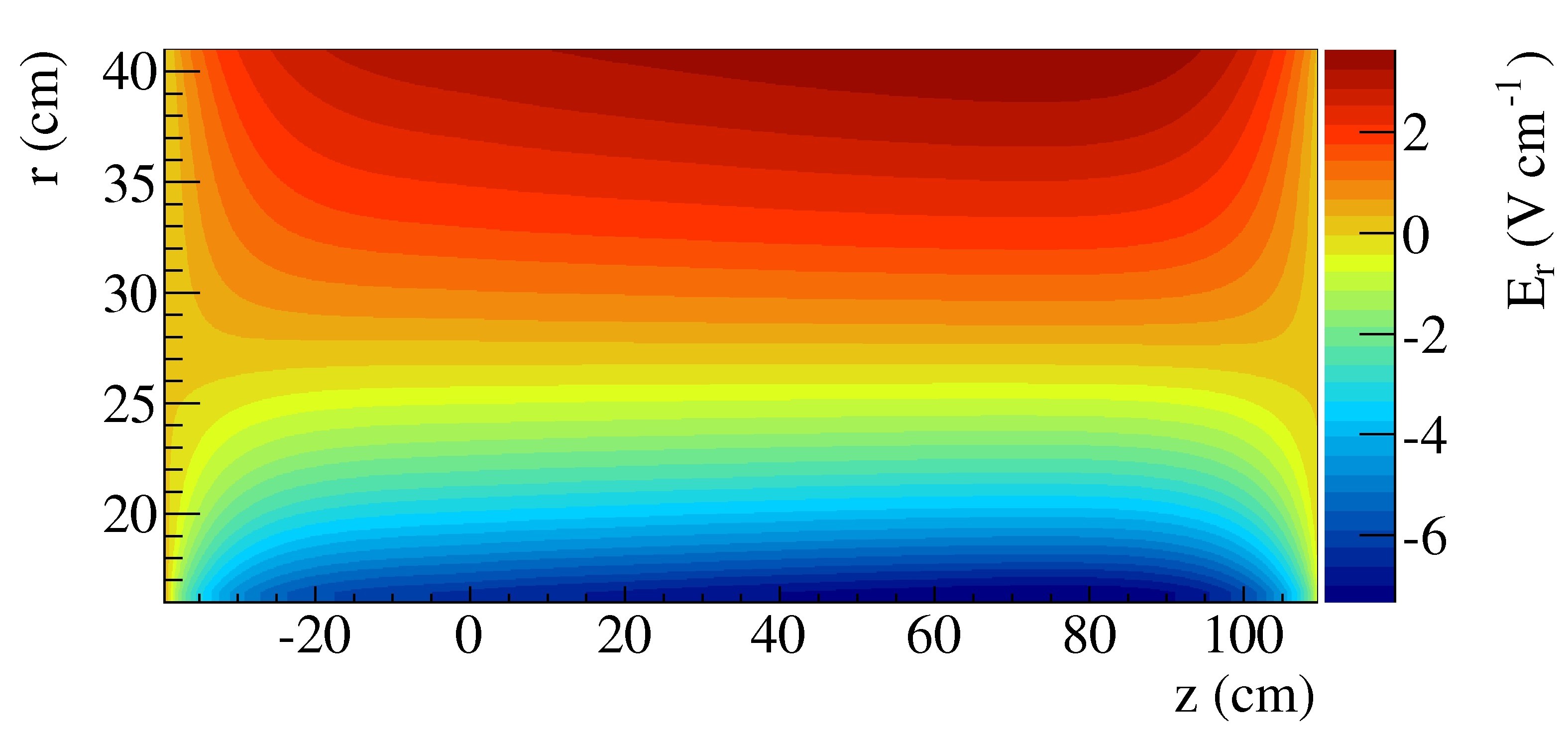}}
  \\ 
  \subfloat[]{\label{fig:efield-2}\includegraphics[width=\linewidth]{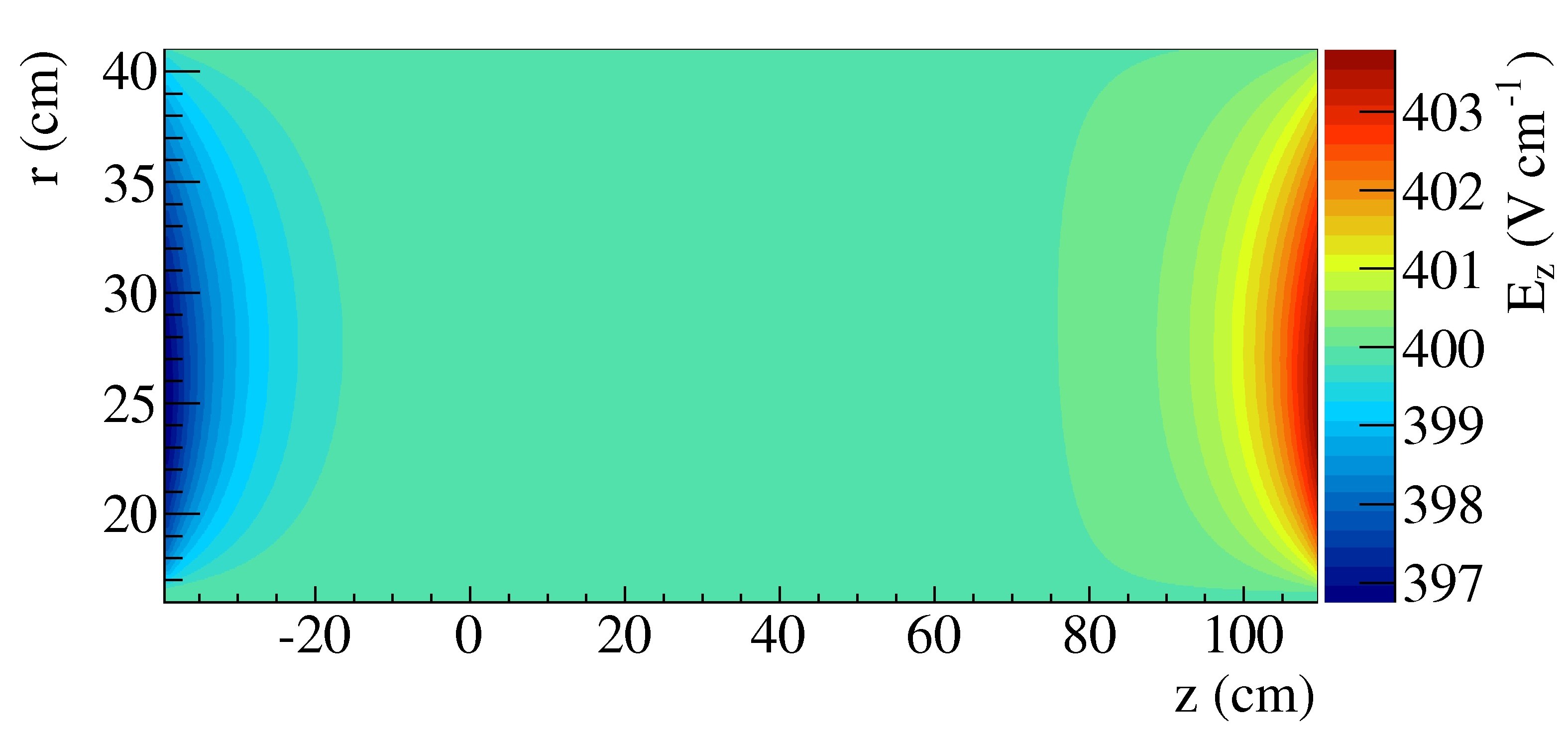}} \\
  \caption{Electric field caused by accumulated space charge in the
    \panda\ TPC for $2\cdot10^7\, \mathrm{s^{-1}}$ $\overline{p}p$
    annihilations at beam momentum $p_{\bar{p}}=2.0\,\GeV/c$ with ion
    back-flow factor $\epsilon=4$. (a) Radial component $E_r$, and (b)
    component along drift direction $E_z$ superimposed with the
    nominal drift field $E_{\text{d}} = 400\,\mathrm{V\,cm^{-1}}$.}
  \label{fig:efield}
\end{figure}

By calculating the gradient of the resulting potential we finally
obtain the electric field $\mathbf{E}$ caused by the space charge. Due
to the assumed $\varphi$-symmetry, the non-vanishing components
of $\mathbf{E}$ are the $r$ and $z$ components.  The drift field
$E_{\text{d}} = 400\,\mathrm{V\,cm^{-1}}$ is oriented along the $z$
direction and is superimposed onto the distortion field (see
Fig.~\ref{fig:efield}).

\subsection{Electron Drift Offsets} 
\begin{figure}[tb]
  \centering 
  \subfloat[] {\label{fig:dev_EB_rad}
    \includegraphics[width=\linewidth]{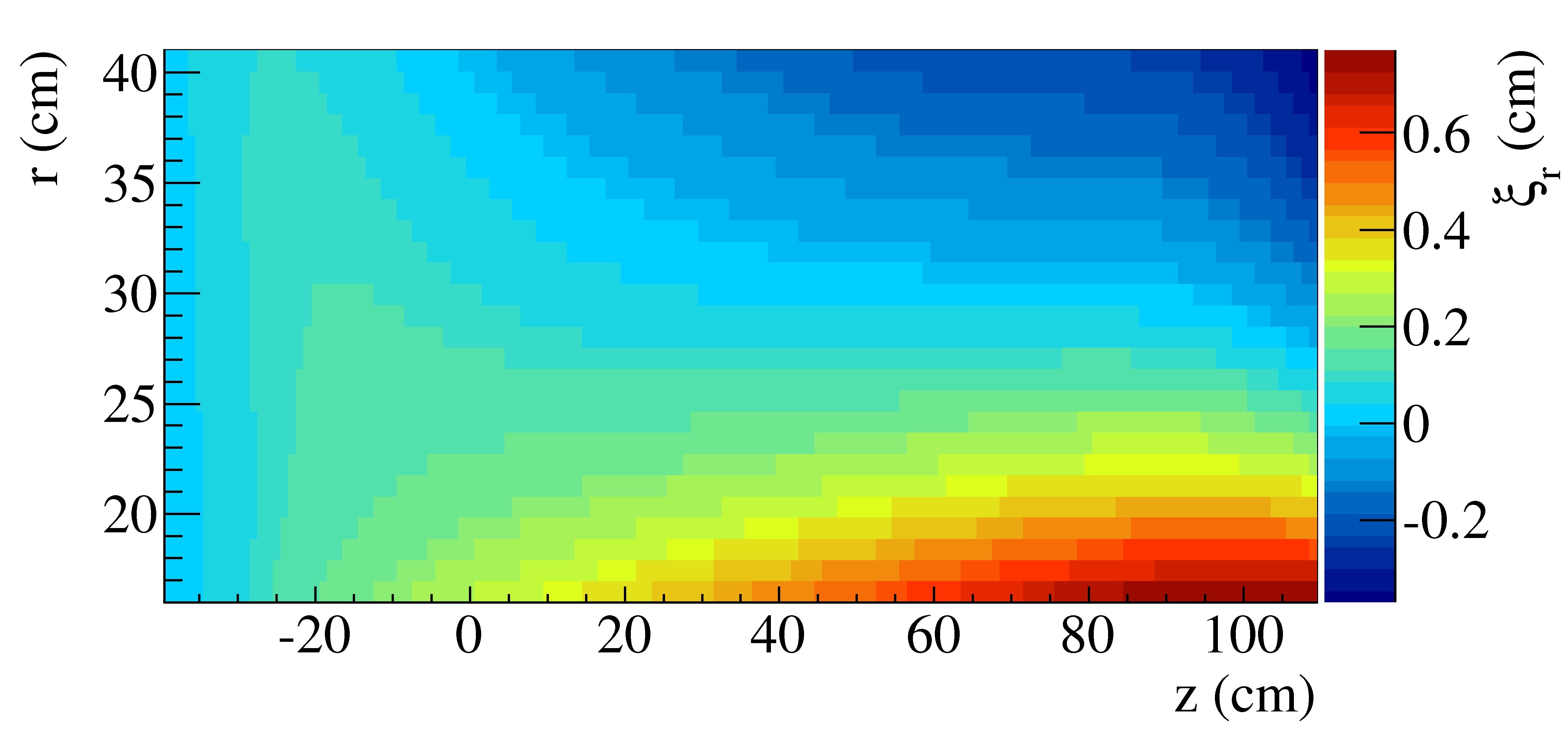}} \\
  \subfloat[] {\label{fig:dev_EB_perp}
    \includegraphics[width=\linewidth]{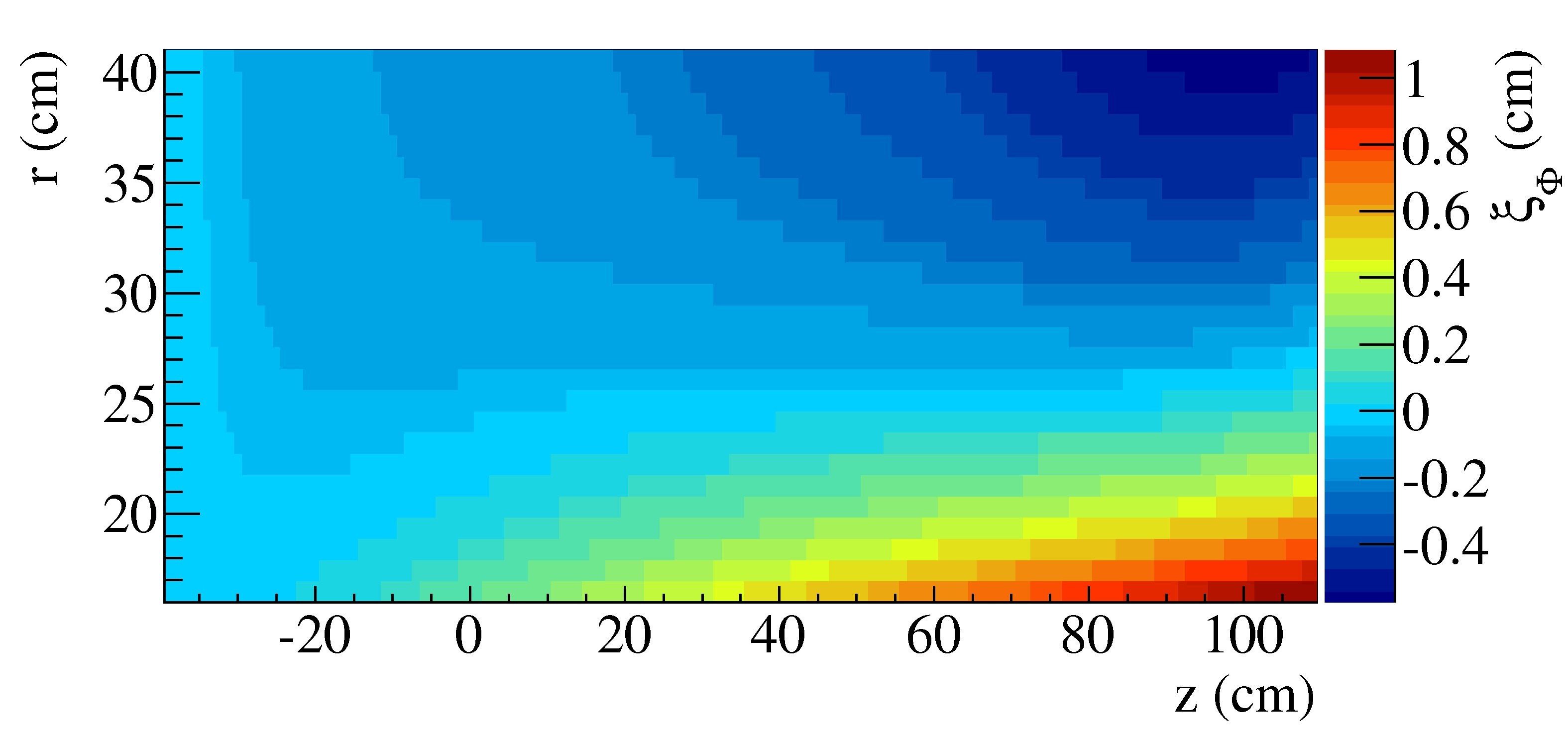}}
  \caption{Drift distortions $\xi$ of electrons due to ion space
    charge accumulation, (a) in radial direction, and (b) in azimuthal
    direction. The graphs show the deviations from a straight line
    drift experienced by an electron which starts its drift at points
    $(r,z)$ in the TPC. The $z$-component of this \textit{distortion
      map} is found to be negligible in comparison.}
\label{fig:devEB}
\end{figure}

To compute a map of the drift distortions as a function of the starting
point of the drift, the equation of motion for electrons drifting in
electric and magnetic fields ($\mathbf{E},\mathbf{B}$),
\begin{equation}
  \label{eqn:eqnmo}
  m\,\frac{\text{d}}{\text{dt}}\mathbf{u}^{-}=e\mathbf{E}
  + e[\mathbf{u}^-\times\mathbf{B}]-K\mathbf{u}^- \quad,
\end{equation}
is integrated with a fourth order Runge Kutta algorithm\footnote{When
reading field values from maps with finite bin size, linear
interpolation in between bin centers is applied.}. In
Eq.~(\ref{eqn:eqnmo}) $m$ is the electron mass, $\mathbf{u}^{-}$ is
the macroscopic electron drift velocity, $e$ is the electron charge
and $K=e/\mu^-$ is a constant friction term modeling the microscopic
stop-and-go motion of the drifting electron due to collisions with the
gas components (mobility $\mu^-=u^-/E_z$ with $E_z$ being the $z$
component of $\mathbf{E}$).  The electric field $\mathbf{E}$ is the
result of the steps described in Sec.~\ref{sec:distE}, while the
magnetic field $\mathbf{B}$ is given by the calculated field map of
the \panda\ solenoid, with a nominal field of $2\,\T$ and a relative
uniformity of better than $2\%$ everywhere in the active volume of the
TPC \cite{PANDA:2009pt}.  The fact that both $\mathbf{E}$ and
$\mathbf{B}$ have non-zero components perpendicular to $z$ leads to a
displacement of drifting electrons compared to the ideal drift with
constant velocity along $z$. Figure~\ref{fig:devEB} shows the radial
and azimuthal components of this deviation at the readout plane as a
function of the coordinates of the starting point of the drifting
electron.  It should be noted that although both fields are assumed to
exhibit cylindrical symmetry, there are nevertheless drift distortions
perpendicular to the $r$-$z$-plane. This is due to the 
$\mathbf{E}\times\mathbf{B}$ term in the solution to
Eq.~(\ref{eqn:eqnmo}).

%% file: sc_recovery.tex
\section{Measurement and Correction of Drift Distortions} 
\label{sc_laser}
In order to correct track data for space charge effects, the resulting
drift distortions have to be measured during the operation of the TPC.
This can be achieved by creating a well defined pattern of electron
sources in the TPC and analyze its image as measured by the
detector. A comparison with the expected image directly yields the
distortions.  Straight line ionization tracks from UV lasers can
provide such a pattern \cite{Hilke:1986} and have already been used
for calibration in 
other drift chambers (e.g. in STAR \cite{Abele:2003}). To assess the
potential of this method in the case of an ungated TPC, we have
simulated such a laser system.

\subsection{Laser Calibration System Simulation}
\label{sec:laser}
Having found the drift
distortions in $z$-direction to be small, we want to measure the
distortions in the $r$-$\phi$-plane as precisely as possible. Ideally,
this can be done with laser tracks parallel to the drift direction /
detector symmetry axis. We chose a grid of 6 laser beams aligned
equidistantly in radial direction, where the innermost (outermost) laser
is placed $0.5\,\text{cm}$ away from the inner (outer) field cage
wall. One such row of lasers is created every $20^{\circ}$ in
azimuthal direction.
The laser beams are parametrized as straight line tracks with an
average ionization density of $45\,\mathrm{e^{-}}\,\mathrm{cm}^{-1}$
\cite{Abele:2003}, and a Gaussian beam profile ($\sigma = 300 \,
\upmu\mathrm{m}$). 

\begin{figure}[tb]
  \centering
  \includegraphics[width=\linewidth]{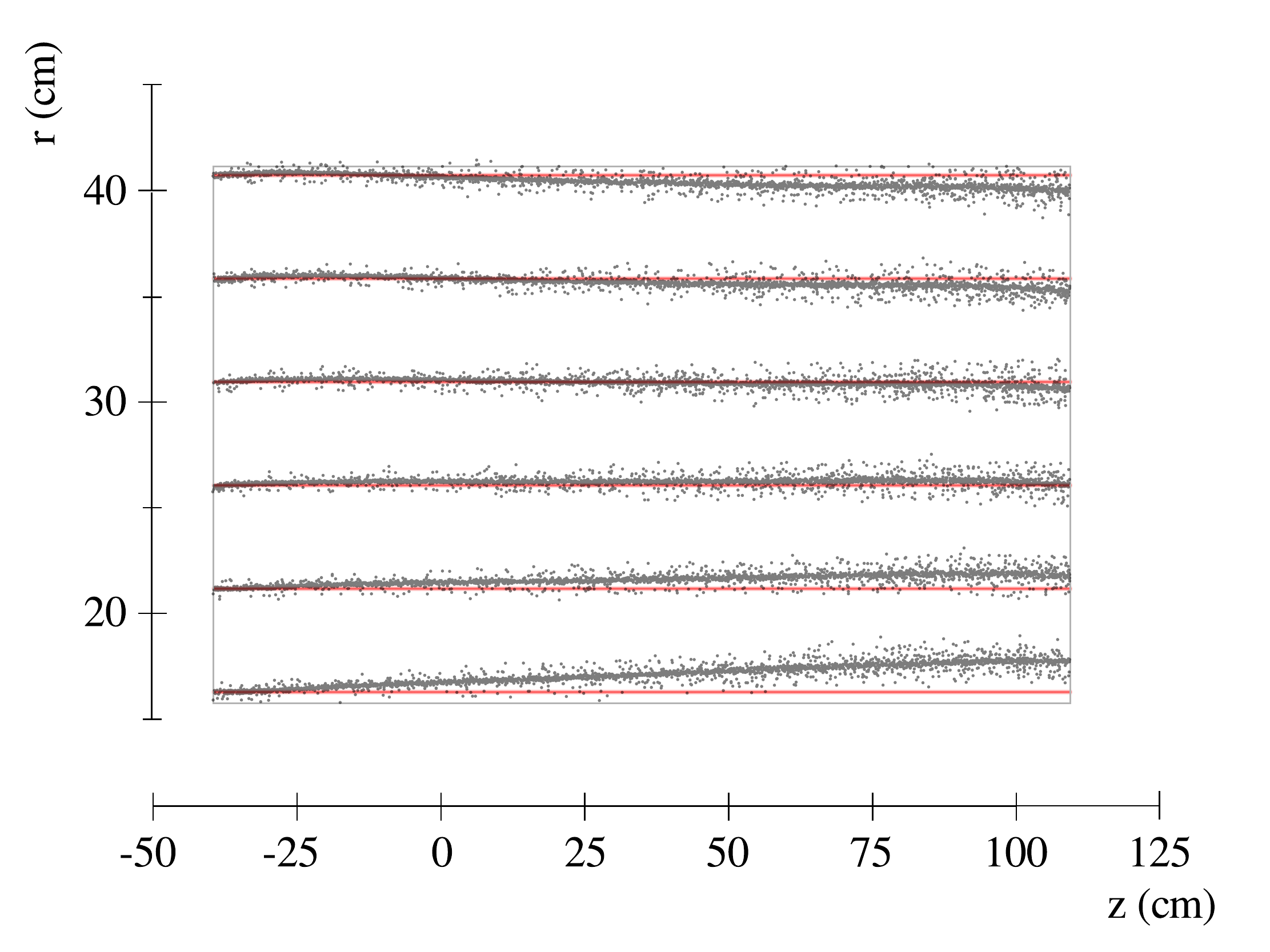}
  \caption{Positions of TPC hits (after running of the clustering
    algorithm) from a laser event.  The laser tracks are sketched as
    lines along the drift direction in the TPC. For better visibility,
    distortions have been scaled by a factor of 2 for this image. Note
    that only the \textit{radial} component of the distortions can be
    seen in this projection.}
  \label{fig:laserResiduals}
\end{figure}

In the simulation routines, the electrons created by the laser rays
are then drifted to the readout anode, applying the calculated
distortion map and taking into account diffusion. The induced signals
are calculated using GEM amplification and signal induction with a
realistic response of hexagonal readout pads with an outer radius of
$1.5\,\mm$ as designed for the \panda~TPC. In a last step, signals
adjacent in space and time are combined to hits by a clustering
algorithm. A sample of such hit data as written out by the simulation
framework is shown in Fig. \ref{fig:laserResiduals}.

\subsection{Reconstruction of  Drift Distortions}
\label{sec:grid}
The reconstruction of the laser tracks is greatly simplified by the
knowledge of the geometry (and the time, at which the laser grid has
been illuminated): The hits are assigned to tracks by applying simple
residual cuts, such that all hits inside a tube of reasonable radius
($15\,\mm$ for the results presented) around the nominal track
position are associated to the corresponding laser track. It is
assured that each reconstructed hit is assigned to only one (the
closest) track.  For each hit, a residual vector is obtained in the
plane perpendicular to the corresponding track.

In order to map out drift distortions as a function of the point of
electron creation in the drift volume, the large amount of signal data
from the laser calibration needs to be fitted, smoothed and
parametrized.  For this purpose we have implemented an algorithm for
fitting of a bi-cubic spline surface $s \mathrm{(}z,r\mathrm{)}$
\cite{HAYES01081974} to the measured residual data (c.f. Fig.
\ref{fig:laserResiduals}).  A two-dimensional mesh of $h \times k$
knots $(\lambda_{i}, \mu_{j})$ over the data area $(z, r)$ is chosen
\footnote{In addition 4 knots on each side outside the data area are
  required to define the full set of B-splines.}. At each of these
knots two cubic B-splines $M_{i}(z)$, $N_{j}(r)$ are attached, each of
which has a fixed shape and spans over five mesh knots
$\lambda_{i-4,\,...,\,i}$ ($\mu_{j-4,\,...,\,j}$). Coefficients
$\gamma_{ij}$ at every knot scale the B-splines:
\begin{equation}
  \label{eq:spline}
  s \mathrm{(}z,r\mathrm{)} = \sum\limits_{i=1}^{h+4}
  \sum\limits_{j=1}^{k+4} \gamma_{ij} \, M_{i}
  \mathrm{(}z\mathrm{)} \, N_{j}\mathrm{(}r\mathrm{)}\quad.
\end{equation}
Assuming the underlying noise (mainly given by diffusion during drift)
to be Gaussian, the surface fitting problem is equivalent to finding the set
of coefficients $\gamma_{ij}$ which minimize the sum of squared
distances 
\begin{equation}
  \label{eq:LS}
S = \sum_{r=1}^{n_{r}} \left[\,s(z_{r}, r_{{r}}) - f_{{r}}\,\right]^2 \equiv \sum_{{r}=1}^{n_{{r}}} R_{{r}}^2
\end{equation}
to the residual data $f_{r} (z_{r},
r_{r})$ (where $r = 1 \,...  \,n_{r}$). The
corresponding matrix-equation (``normal equation'') reads:
\begin{equation}
  \label{eq:matrix}
\mathbf{A}^T \mathbf{A \, \bm{\gamma}} = \mathbf{A}^T \mathbf{f}\quad,
\end{equation}
where $\mathbf{A}$ is an $n_{r} \times
\mathrm{(}h+4\mathrm{)(}k+4\mathrm{)} $-dimensional matrix containing
the spline information, \boldmath$\gamma$ \unboldmath is the vector of knot
coefficients with $\mathrm{(}h+4\mathrm{)(}k+4\mathrm{)}$ elements and
$\mathbf{f}$ is the vector containing the $n_{r}$ measured residuals.
Weighting of hits can be incorporated by introducing a weight matrix
$\mathbf{W}$. Under the assumption of independent hit errors,
$\mathbf{W}$ becomes diagonal, and instead of Eq.~(\ref{eq:LS}) one
now has to minimize
\begin{equation}
  \label{eq:LS_weight}
S_{\text{w}} =  \sum_{r=1}^{n_{{r}}} \text{W}_{{rr}}R_{{r}}^2\quad.
\end{equation}
Equation~(\ref{eq:matrix}) then reads
\begin{equation}
  \label{eq:matrix_weight}
\mathbf{A^{\prime}}^T \mathbf{A^{\prime} \, \bm{\gamma}} =
\mathbf{A^{\prime}}^T \mathbf{f^{\prime}}\quad,\end{equation} where
$\mathbf{A^{\prime}} = \mathbf{w}\mathbf{A}$, $\mathbf{f^{\prime}} =
\mathbf{w}\mathbf{f}$, and $\text{w}_{{rr}} =
\sqrt{\text{W}_{{rr}}}$. The weights for each hit are obtained
from the covariance ellipsoid available after clustering, 
taking into account the 
orientation of the laser track. 
By setting the error to be
large along the track direction, it is possible to work also with
laser grids that are not optimized to the distortions topology.

\begin{figure}[tb]
  \centering 
  \subfloat[]{\label{fig:rec_devmapRad}\includegraphics[width=\linewidth]{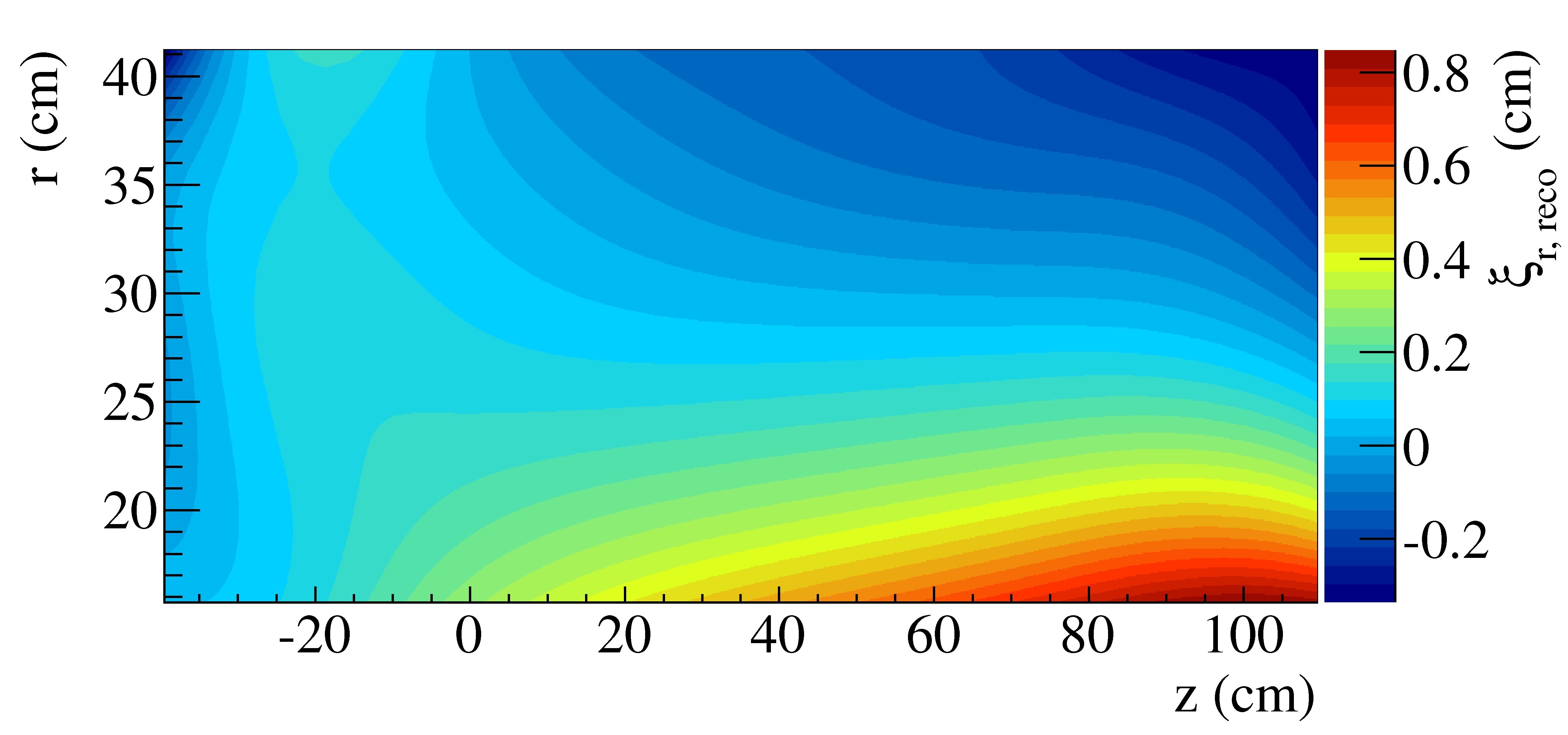}}
  \\
  \subfloat[]{\label{fig:rec_devmapPhi}\includegraphics[width=\linewidth]{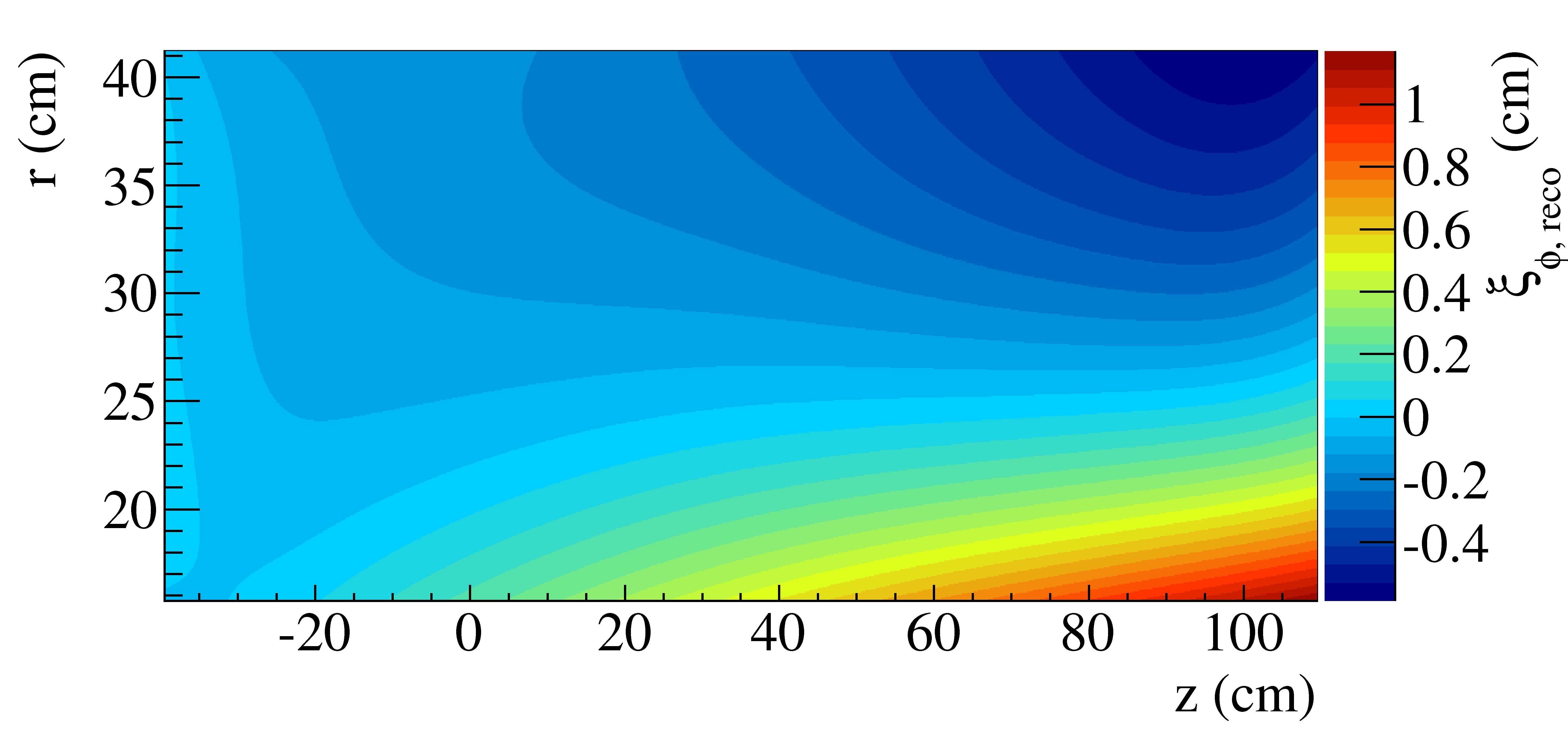}}
  \caption{Reconstructed drift distortions $\xi$ in (a) radial
    direction and (b) azimuthal direction, based on one laser event
    (compare to Figs.~\ref{fig:dev_EB_rad} and \ref{fig:dev_EB_perp}).  }
  \label{fig:rec_devmap}
\end{figure}

The solution to Eq.~(\ref{eq:matrix_weight}) can be obtained by matrix
inversion or, more stably, by QR-decomposition.
Figure~\ref{fig:rec_devmap} shows the least
squares spline solution to the \textit{reconstructed} residual data
for both radial and azimuthal drift distortions. One laser event 
as described in Sec.~\ref{sec:laser} 
is used, resulting in $\sim 35,000$ reconstructed residuals.  
Typical fit times are of the order of $1\,\s$ on a single CPU desktop PC.
  
\subsection{Quality of Reconstructed Distortion Map}
The choice of the number of spline-mesh knots $h\times k$ over the
data area is very important for the procedure.  On the one hand, a
higher number of knots means higher overall fit precision, but
overfitting becomes an issue. A lower number of knots, on the other
hand, results in a smoother result, but the fit might not be able to
follow the important features of the data set. We found $(h\times k) =
5\times 3$ to be a balanced value. 

\begin{figure}[tb]
\begin{center}
\subfloat[]{\label{fig:reco_diffRad}\includegraphics[width=\linewidth]{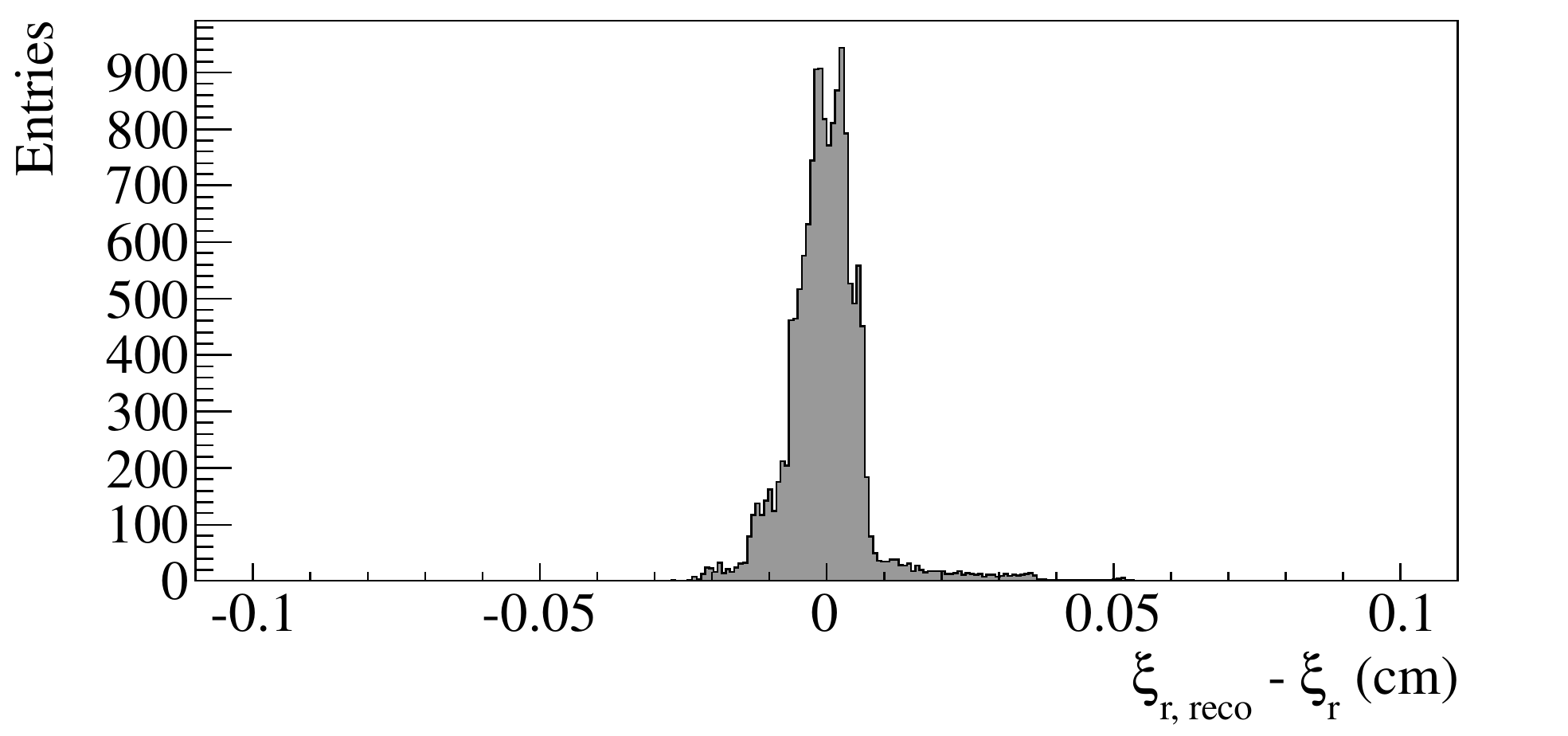}}
\\
\subfloat[]{\label{fig:reco_diffPhi}\includegraphics[width=\linewidth]{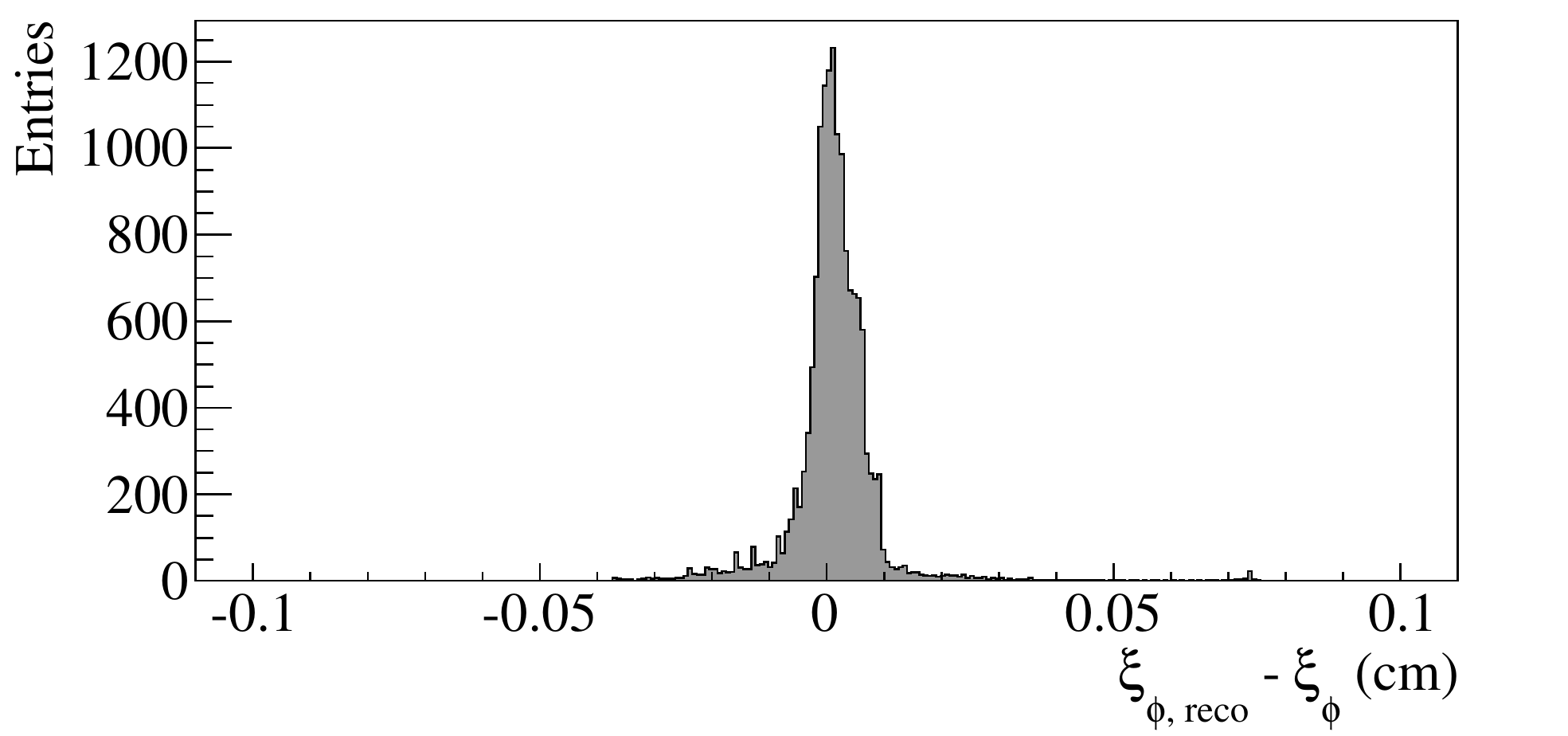}}
\end{center}
\caption{Direct comparison of drift distortions as reconstructed with
  the Laser grid (spline fit) and original simulation input. Shown are
  the differences of the original (binned) deviation map and the fit
  evaluated at the bin center for the (a) radial and (b) azimuthal
  drift distortions. The corresponding r.m.s.~of the histograms are
  $72\,\upmu \text{m}$ for (a) and $86\,\upmu \text{m}$ for (b), respectively.}
\label{fig:reco_diff}
\end{figure}

To be able to judge the quality of
the result shown in Fig.~\ref{fig:rec_devmap}, we compare the
reconstructed distortion maps to our original drift distortion input
(Fig.~\ref{fig:devEB}) in Fig.~\ref{fig:reco_diff}, which shows the
distribution of differences between the reconstructed and the
calculated distortion maps.

Fitting a Gaussian to the peak of the distributions gives
$\sigma_{\mathrm{Gauss}} \lsim 50 \, \upmu\text{m}$ and a mean well
centered around 0, showing that the total systematic error of the
laser correction method is small and the overall precision of the
presented distortion reconstruction and fitting method is better than
the expected single-hit resolution of the TPC. Hence, using a
grid of laser beams, drift distortions in the TPC can be
measured with a precision better than $\mathcal{O}$($\mathrm{100 \,
  \upmu m})$, not taking into account mechanical distortions of the
calibration system.

\subsection{Effect on Track Reconstruction}
The most important question to be answered is of course the impact of
space charge effects on track 
reconstruction.  As explained in the previous sections, the steps
required to obtain a model of space charge and study its possible
effects and their correction are:
\begin{enumerate} 
 \item calculate space charge based on background events; 
 \item calculate the drift distortion map;
 \item create one ``laser event'' consisting of the full laser beam mesh; 
 \item reconstruct the laser tracks and fit the data with a spline; 
 \item use this spline to correct for drift distortions.
\end{enumerate}
In the running experiment, measurements of drift distortions have to be
constantly updated, requiring steps 4 and 5 to be triggered on a
timescale short enough to be able to resolve possible space charge
fluctuations.

\begin{figure}[tb]
\centering
\includegraphics[width=\linewidth]{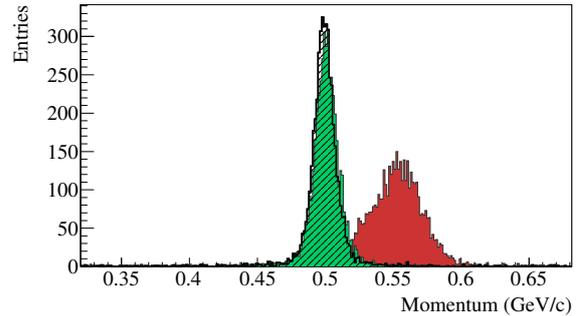}
\caption [Effect and Correction of drift Distortion in Reconstructed
Tracks]{Effect of drift distortions (and their correction) for
  $\pi^{+}$-tracks at $0.5 \, \text{GeV}/c$ momentum (primary and
  secondary GEANT tracks). Right peak: Uncorrected case. Left peak:
  Hit positions corrected with spline fir results. Transparent
  distribution: Simulation with no drift distortions present. }
\label{fig:dev_correction}
\end{figure}

To study the effect of drift distortions on charged particle tracks,
we simulated a test sample of 5000 pion tracks ($\pi^{+}$) at a
momentum of $0.5 \, \text{GeV}/c$ and uniformly distributed
scattering angle ($20^{\circ} \leq \theta \leq 115^{\circ}$) in the
\panda~TPC. 
Tracks are reconstructed using a Kalman-filter-based track fitting
framework \cite{Hoeppner:2010a}.  
Figure~\ref{fig:dev_correction}
visualizes the impact of 
uncorrected drift distortions on the momentum measurement compared to the
ideal situation of a completely homogeneous electrical field.
Without correction, the distribution
of reconstructed momenta is deformed and systematically shifted
towards larger values by space charge effects. This can be understood
within the scope of the preceding sections: Depending on its polar
angle, a track is more or less affected by drift distortions (c.f.\
Fig.~\ref{fig:devEB}), leading to the broadening of the
distribution. The asymmetric nature of the drift distortions with
respect to the radial coordinate causes the measured curvature of the
track to appear smaller than it actually was, leading to the shift to
higher momenta.  Applying the correction derived from the laser system
essentially recovers the momentum reconstruction, with a resolution of
$\sigma_p/p=1.74\,\%$.  For comparison, the distribution of
reconstructed momenta for the ideal case without any space charge
effects yielding a resolution of $\sigma_p/p=1.59\,\%$ is shown as
outline in Fig. \ref{fig:dev_correction}. The reconstructed mean
changes from $5.01\cdot 10^{-1}\,\text{GeV}/c$ in the ideal case to
$4.99\cdot 10^{-1}\,\text{GeV}/c$, respectively.

%% file: sc_conclusion.tex
\section{Conclusions and Outlook}
In this paper we have presented detailed studies on space-charge
buildup in a continuously operating GEM-TPC without a gating grid
exposed to high interaction rates. 
We have presented a recursive algorithm to calculate the space charge 
accumulating in the drift region, starting from minimum bias physics
events, 
and taking into account the primary ionization as well as the ions
drifting back from the amplification region. 
Assuming constant luminosity, an equilibrium space-charge 
distribution is reached after one full drift time of ions through the
chamber. 
In the example of the
environment of the future \panda~detector,
ion space charges of up to $65 \mathrm{ \, fC \, cm^{-3}}$ are
reached. 
The electric field resulting from this space charge is
calculated using a finite element method to solve the Poisson
equation. The drift of electrons in the full electric and magnetic
field of the setup is calculated from the Langevin equation solved by
a fourth order Runge-Kutta method. Drift-path distortions of
$\mathcal{O}$($1 \, \mathrm{cm}$) are found for the \panda\
enviromment, requiring  
corrections to be applied to the measured hit coordinates.

To this end, we have simulated a laser calibration system creating a
regular grid of straight ionization tracks at known positions in the
chamber. The geometry of the laser grid was optimized to yield an
accurate measurement of the expected distortions, which are dominantly
in radial and azimuthal direction for the \panda\ case.  
In the case of azimuthal symmetry of the distortions, a
two-dimensional spline fit to the measured residuals is sufficiently
fast and accurate, reproducing distortions with a precision better
than $100\,\upmu\m$, thus opening the possibility to be applied even
in the online reconstruction software.  This distortion map is then
used to correct hits from charged particle tracks. Using a
Kalman-Filter-based track fitting we are able to reconstruct the
momentum of tracks with a resolution very close the ideal case of no
distortions in the chamber.

The method to calculate the space charge effects and their correction
presented in this paper is also applicable to the case of 
other TPCs with azimuthal symmetry, e.g.\ at STAR or ALICE. 
If one cannot exploit symmetry features of the distortions 
or it is not possible to realize the optimal laser grid geometry due
to technical constraints, proper treatment of errors taking into
account the grid geometry, as realized in the present work, is
important to minimize systematic errors 
arising from residual projections. 
This allows the algorithm to be 
extended to make use of track information from other detectors instead
of a fixed laser grid to measure the drift distortions.